\documentclass[aip,jap,preprint]{revtex4-1}

\usepackage{epsfig}
\usepackage{epstopdf} 

\usepackage{subfig}

\usepackage{titlesec}
\titleformat{\section}[hang]
  {\centering}{\thesection}{1ex}{\normalsize \textsc}
\titleformat{\subsection}[hang]
  {}{\thesubsection}{1ex}{\normalsize \textit}

\renewcommand{\thesection}{ \normalsize \textnormal{\Roman{section}.}}
\renewcommand{\thesubsection}{\normalsize \textnormal{\textsc{\textit{\Alph{subsection}.}}}}

\def\e{\begin{equation}}
\def\f{\end{equation}}
\def\_#1{{\bf #1}}

\def\.{\cdot}

\newcommand{\epsdyad}{\ensuremath{\overline{\overline{\epsilon}}}}
\newcommand{\mudyad}{\ensuremath{\overline{\overline{\mu}}}}

\begin{document}

%
\author{\normalsize \bfseries J. Vehmas,$\mathbf{^1}$ S. Hrabar,$\mathbf{^2}$ and S. Tretyakov}
\affiliation{$\mathbf{^1}$ Department of Radio Science and Engineering/SMARAD Center of Excellence, Aalto University, P.~O.~Box~13000, FI-00076 Aalto, Finland. \\
Email: joni.vehmas@aalto.fi\\ $\mathbf{^2}$ Faculty of Electrical Engineering and Computing, University of Zagreb, Unska 3, 10000 Zagreb, Croatia.
}
%
%
%
\title{\large \textbf{Omega transmission lines with applications to effective medium models of metamaterials}}


\begin{abstract}
In this paper we introduce the concept of transmission lines with
inherent bi-anisotropy and establish an analogy between these lines
and volumetric bi-anisotropic materials. In particular, we find
under what conditions a periodically loaded transmission line can be
treated as an effective omega medium. Two example circuits are
introduced and analyzed. The results have two-fold implications:
opening a route to emulate electromagnetic properties of
bi-anisotropic omega media using transmission-line meshes and
understanding and improving effective medium models of composite
materials with the use of effective circuit models of unit cells.
\end{abstract}
\maketitle

\section{Introduction}

Omega media, introduced in Refs. \onlinecite{Saadoun} and \onlinecite{proposed}, is a
widely researched reciprocal special case of bi-anisotropic media.
In bi-anisotropic media, electric field induces both electric and
magnetic polarizations (and the same is true for magnetic field).
The omega material is reciprocal and it is characterized by an
antisymmetric magnetoelectric coupling dyadic. In case of the
uniaxial omega media \cite{proposed} the material relations can be
written as
\begin{equation}
\begin{array}{l}
\mathbf{D}=\epsdyad \cdot \mathbf{E} + j K \sqrt{\epsilon_0 \mu_0}\, \mathbf{z}_0 \times \mathbf{H}
\\
\mathbf{B}=\mudyad \cdot \mathbf{H} + j K \sqrt{\epsilon_0 \mu_0}\, \mathbf{z}_0 \times \mathbf{E}.
\end{array} \label{eq:omegamatrel}
\end{equation}
%
Here, $\mathbf{E}$ and $\mathbf{H}$ are electric and magnetic fields while $\mathbf{D}$ and $\mathbf{B}$ stand for electric and magnetic flux densities, respectively. Furthermore, $K$ is the omega coupling coefficient, $\epsilon_0$ and $\mu_0$ are, respectively, the vacuum permittivity and permeability, and the unit vector $\_z_0$ defines the only preferred
direction in the uniaxial sample. The permittivity and permeability
dyadics (denoted as $\epsdyad$ and $\mudyad$) are symmetric and uniaxial.

Omega materials have an interesting property of having different
wave impedances for opposite propagation directions. This implies
asymmetry of reflection coefficients at interfaces with omega media
when the illumination direction is reversed. Furthermore, the
matching condition for interfaces between omega media and conventional magnetodielectric media depends on all
three material parameters, opening interesting design possibilities
in antenna and microwave engineering \cite{proposed,Serdyukov}. Most
of recently introduced terahertz and optical metamaterial structures
have the symmetry corresponding to the omega type of bi-anisotropic
coupling (for example, arrays of complex-shaped metal particles or
meshes positioned on one side of a dielectric substrate or similar
multi-layer structures). Understanding and modeling of
magnetoelectric omega coupling  is a pre-requisite for understanding
of effective response of these advanced electromagnetic materials
and developing new applications.

It has been suggested that omega media can be realized by embedding
electrically small resonant metal particles of an appropriate shape
(e.g., $\Omega$-shaped metal inclusions, that is, centrally connected
small dipole and loop antennas), into a conventional dielectric \cite{Saadoun}.
However, as such wire omega particles are resonant structures with
all the polarizabilities (electric, magnetic, magneto-electric, and
electro-magnetic) resonant always at the same frequency, the
operational bandwidth, i.e., the bandwidth where $K$ is
significantly different from zero, is rather narrow. This also means
that the tunability of medium parameters in general is very limited
and such media always have band-stop behavior. Losses in such media are typically also quite high when $K$ is significantly large. For these reasons, the possibilities for practical applications are limited. Here, we consider
possibilities of realizing omega media with periodically loaded
transmission lines (TLs). First, we compare the wave impedance of
omega media with the Bloch impedance of a general periodically
loaded TL and derive the required conditions for omega-like
response. Second, a T-type circuit topology is considered to fulfill
the required conditions. Third, a circuit topology using coupled
inductors is analyzed. It should be emphasized that the goal, here, is not to replicate the narrow-band dispersion of any of the proposed realizations based on resonant particles, but rather to study more general media that still satisfy the material relations for omega media.

As will be shown below, the  omega coupling parameter is defined by
the asymmetry of unit cells of periodical structures (transmission
lines in this example). This has important implications for problems
of homogenization of composite materials (metamaterials) and
electrically thin composite layers (metasurfaces). The unit cells
(or periodically repeated planes of various inclusions) can be
modeled by equivalent T- or $\Pi$-circuits, arranged in a periodical
fashion. Here we will show that the asymmetry of these unit cells
(most commonly imposed by the topology of the sample interfaces) can
be properly accounted for by the effective omega-coupling parameter.
This can remove the common problem of non-physical anti-resonance in
effective permittivity and permeability, extracted from the
reflection and transmission coefficients of planar slabs, without
the need to introduce additional parameters which explicitly depend
on the propagation constants of partial plane waves in the medium
(as in Refs. \onlinecite{alu1,alu2}). From the physical point of view, the
effective omega parameter accounts for first-order spatial
dispersion effects in materials with non-negligible electrical size
of the unit cells.

\section{Omega media and omega transmission lines}

\subsection{Propagation constant and wave impedance}

The propagation constant for axially propagating  plane waves in
omega media can be easily derived from (\ref{eq:omegamatrel}) and is
given by \cite{proposed,Serdyukov}
\begin{equation}
\beta = k_0 \sqrt{\epsilon_t \mu_t - K^2} = k_0 \sqrt{\epsilon_t \mu_t}\sqrt{1 - K_n^2},
\label{eq:omegadispersion}
\end{equation}
where $k_0$ is the free-space wave number, $\epsilon_t$ and $\mu_t$ are, respectively, the relative
transverse permittivity and permeability, and $K_n$ is the normalized
omega coefficient defined as $K_n = K/\sqrt{\epsilon_t \mu_{t}}$.
For lossless media with $\epsilon_t \mu_t > 0$, $K_n$ is purely
real. Therefore, for such media the propagation constant is real,
i.e., there is wave propagation only when we have $|K_n| < 1$. An
interesting property that separates omega media from conventional
magnetodielectric  media is that the wave impedance is different for
waves traveling in the opposite directions. The wave impedance for
the axial propagation can be written as \cite{proposed,Serdyukov}
\begin{equation}
Z_\Omega = \sqrt{\frac{\mu_0 \mu_t}{\epsilon_0 \epsilon_t}}\Big(\sqrt{1-K_n^2} \pm j K_n\Big),
\label{eq:omegaimp_normal}
\end{equation}
where the two solutions correspond to opposite axial propagation
directions.


\subsection{Required conditions for unit cells}
Our goal will be to emulate the wave-propagation properties of omega
media with periodically loaded transmission lines. Bloch impedance
can be considered as the characteristic impedance of periodically
loaded transmission lines. It is defined simply as the ratio of the
voltage and current at the terminals of the unit cell. It should be
noted that the value of the Bloch impedance depends on how the
terminal points are chosen and is, therefore, not unique for a given
unit cell. The Bloch impedance for a general reciprocal periodic
structure is defined using ABCD-parameters of unit cells as
\cite{Pozar}
\begin{equation}
Z_B = \mp \frac{2 B}{A-D\mp\sqrt{(A+D)^2-4}}.
\label{eq:Bloch1}
\end{equation}
Here, the two signs correspond to different propagation  directions
and the current is defined to flow always in the direction of the
energy propagation. It should be noted that the top sign does not
necessarily always lead to the correct solution for the positively
traveling wave and the bottom sign for the negatively traveling
wave, but the solutions may switch. Incorrect choice of sign leads to the non-physical result of negative
real part of the Bloch impedance for passive structures. Taking this
into account, the Bloch impedance for waves traveling along the positive
($+$) and negative ($-$) directions can be written as
\begin{equation}
Z_{B\pm} = \frac{j B}{AD-1}\left(\sqrt{1-\left(\frac{A+D}{2}\right)^2}\pm j \frac{D-A}{2} \right).
\label{eq:Bloch3}
\end{equation}

%
%
%
Comparing the two impedances of (\ref{eq:Bloch3}) to the
wave impedances of omega media (\ref{eq:omegaimp_normal})
and assuming them to be equal, we can write the normalized omega
coefficient $K_n$ in terms of the Bloch impedances $Z_{B+}$ and
$Z_{B-}$:
\begin{equation}
K_n= \frac{\frac{Z_{B+}-Z_{B-}}{Z_{B+}+Z_{B-}}}{\sqrt{1-\left(\frac{Z_{B+}-Z_{B-}}{Z_{B+}+Z_{B-}}\right)^2}}.
\label{eq:Kn_cond}
\end{equation}
This can be further written using the ABCD parameters as
\begin{equation}
K_n= \frac{D-A}{2} \frac{1}{\sqrt{1-A D}}.
\label{eq:Kn_cond2}
\end{equation}
Therefore, as long as we have $A \neq D$, i.e, the unit cell is
asymmetric, and $AD$ is finite, the normalized omega coefficient
$K_n$ is non-zero.
Furthermore, we can also determine the effective magnetodielectric wave impedance $\sqrt{\mu_t \mu_0/( \epsilon_t \epsilon_0)}$ based on (\ref{eq:omegaimp_normal}) and (\ref{eq:Bloch3}). This can be written as
\begin{equation}
\sqrt{\frac{\mu_0 \mu_t}{\epsilon_0 \epsilon_t}} = - \frac{j B}{\sqrt{1-A D}}.
\end{equation}
Knowing the effective normalized omega coefficient and the effective
wave impedance,  we can also extract the effective refractive index
by comparing the dispersion in omega media
(\ref{eq:omegadispersion}) and the dispersion in the periodically
loaded TL. The latter can be calculated easily for any unit cell
using basic ABCD-matrix theory and the Floquet theorem and is given
by \cite{Pozar}
\begin{equation}
\beta_\pm = \mp  \frac{j}{d} \ln \left(\frac{A+D\pm \sqrt{(A+D)^2-4(A D-B C)}}{2}\right),
\label{eq:dispersion1}
\end{equation}
where $d$ is the period of the structure. For reciprocal unit cells ($A D - B C = 1$), the two solutions are equal and (\ref{eq:dispersion1}) simplifies to
\begin{equation}
\beta = -\frac{j}{d} \ln \left(\frac{A+D}{2} + \sqrt{\left(\frac{A+D}{2}\right)^2-1}\right).
\label{eq:dispersion2}
\end{equation}
Knowing the effective  normalized omega coefficient, the effective
refractive index of the TL can be determined by comparing
(\ref{eq:dispersion2}) to (\ref{eq:omegadispersion}). Moreover,
knowing the effective refractive index and wave impedance we can
easily find the effective permittivity, permeability, and the
(denormalized) omega coefficient.

\subsection{T circuit}
\begin{figure}[h!]
\centering \epsfig{file=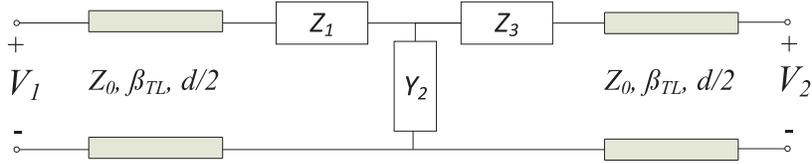, trim = .1cm .0cm .1cm .1cm, clip = true, width=0.65\textwidth}
\caption{The unit cell under study.} \label{fig:unitcell2}
\end{figure}

The simplest possible TL loading element is a T-type circuit (or
alternatively $\Pi$-type), as we need an asymmetric circuit for the
Bloch impedances for different propagation directions to be
different. Let us take a look at a TL periodically loaded with
T-type circuits as shown in Fig.~\ref{fig:unitcell2}. Let us also
assume that the period of the unit cell $d$ is very small
electrically. In this case, the ABCD-parameters, calculated by
simply multiplying the ABCD matrices of each element in the unit
cell in the right order, have the form

\begin{equation}
A = 1+ Y_2{Z_1}, \ \ \ B = Z_1 + Z_3 + Y_2 Z_1 Z_3, \ \ \ C = Y_2, \ \ \ D = 1+ Y_2{Z_3}.
\end{equation}
Therefore, the normalized omega coefficient, as defined in (7), can be written as
\begin{equation}
K_n =  \frac{Y_2 (Z_3 - Z_1)}{2 \sqrt{1-(1+ Y_2{Z_1})(1+ Y_2{Z_3})}}
\end{equation}
and the magnetodielectric wave impedance as
\begin{equation}
\sqrt{\frac{\mu_0 \mu_t}{\epsilon_0 \epsilon_t}} = - \frac{j (Z_1 + Z_3 + Y_2 Z_1 Z_3)}{\sqrt{1-(1+ Y_2{Z_1})(1+ Y_2{Z_3})}}.
\label{eq:mu_over_eps}
\end{equation}
Again, we can see that for a symmetrical unit cell ($Z_1 = Z_3$) we have $K_n=0$.

\begin{figure}[!ht]
  \centering
  \subfloat{\label{fig:comparisona}\includegraphics[width=0.7\textwidth]{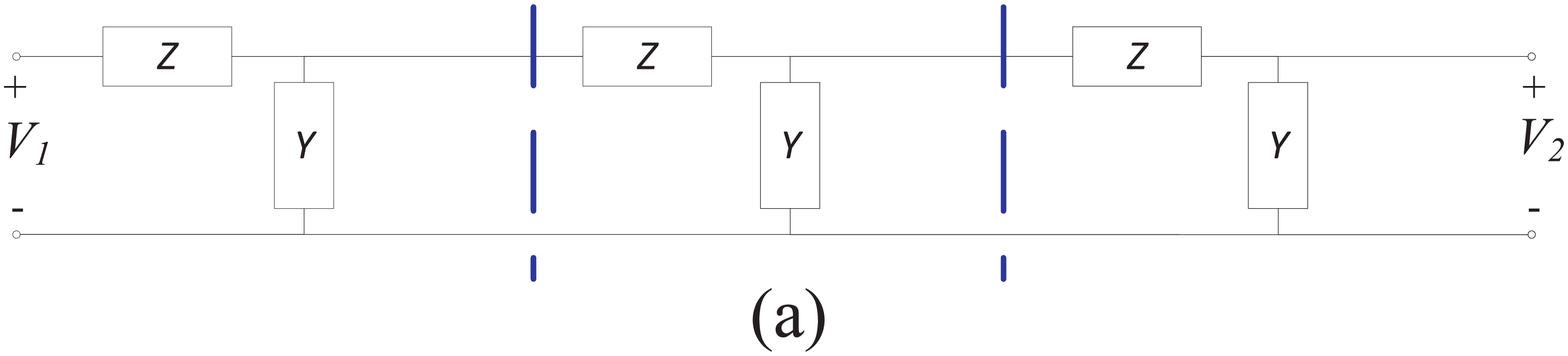}} \vspace{0.5cm}
  \subfloat{\label{fig:comparisonb}\includegraphics[width=0.7\textwidth]{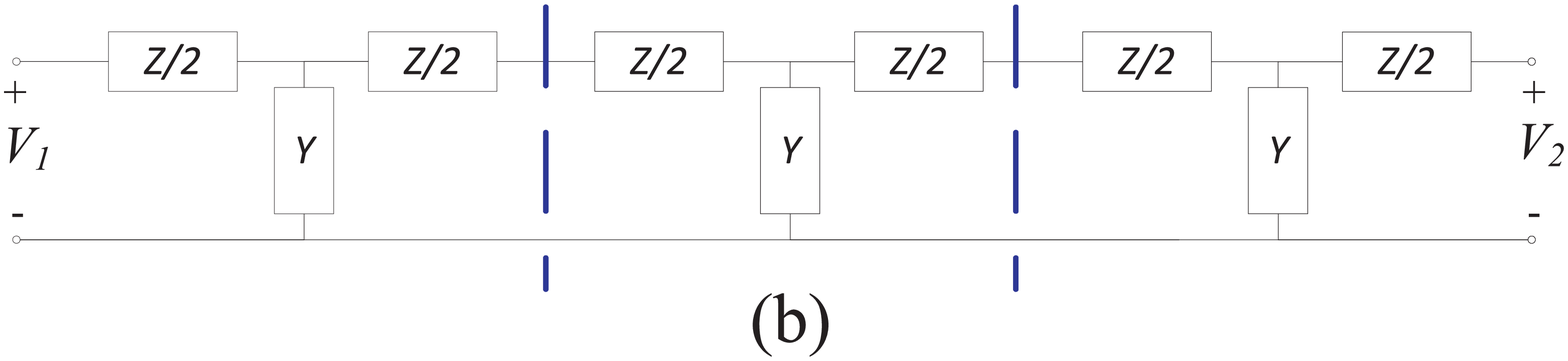}}
  \caption{By defining the terminal points in different ways, the unit cell in an infinite cascade can be asymmetric (a) or symmetric (b).
  Though dispersion is the same for the two structures, the Bloch impedance and therefore also the effective material parameters are different.}
  \label{fig:comparison}
\end{figure}

Let us further simplify the analysis by assuming that we have $Z_3=0
\ \Omega$, that is, the structure corresponds to the cascade shown in Fig.~\ref{fig:comparisona}. In this case, we have simply
\begin{equation}
K_n = \frac{ \sqrt{- Y_2{Z_1}}}{2}, \ \ \ \ \ \ \ \sqrt{\frac{\mu_0 \mu_t}{\epsilon_0 \epsilon_t}} =  \sqrt{\frac{Z_1}{Y_2}}.
\label{eq:LCpars}
\end{equation}
From (\ref{eq:LCpars}), rather surprisingly, it can be seen that
even a TL periodically loaded with a simple two-port consisting of a series inductor and a shunt
capacitor can be interpreted as an omega media with $K_n =
\omega \sqrt{L C}/2$ and $\sqrt{\mu_0 \mu_t / (\epsilon_0
\epsilon_t)} = \sqrt{{L}/{C}}$. Notably, if we would have chosen
$Z_1$ to be zero instead of $Z_3$, the magnetodielectric wave
impedance would be the same but the normalized omega coefficient
would have a different sign ($K_n = -\omega \sqrt{L C}/2$). However,
it is well known that the equivalent circuit of an infinitesimal
section of an unloaded infinite TL can also be considered as a two-port consisting of
a series inductor and a shunt capacitor. Therefore, loading a regular
TL with these lumped elements is typically believed to just increase the
equivalent permittivity and permeability with no need for any extra
effective parameters. In fact, if we define the terminal points of
the unit cell so that the unit cell is symmetric (i.e.,
series-shunt-series loading with  the element values $Z/2$, $Y$,
and $Z/2$ as shown in Fig.~\ref{fig:comparisonb}), we have $D=A$ and thus, according to (7), $K_n = 0$ at
all frequencies. As we consider here an infinite cascade, both unit
cell choices illustrated in Fig.~\ref{fig:comparison} are equally
valid. Obviously, the dispersion in both cases is the same as the
physical structure is the same (assuming that $d$ is electrically
small). On the other hand, looking at (2) the different values of $K_n$ in these two cases indicate that the
effective permittivity and permeability should be defined
differently! In the former case, the stopband is
interpreted to appear due to having $|K_n| > 1$ whereas in the
latter case it appears due to the term $\sqrt{\epsilon_t \mu_t}$.
Let us look at an example in order to clarify this confusion.


%
%
%
%

\begin{figure}[!ht]
\centering \epsfig{file=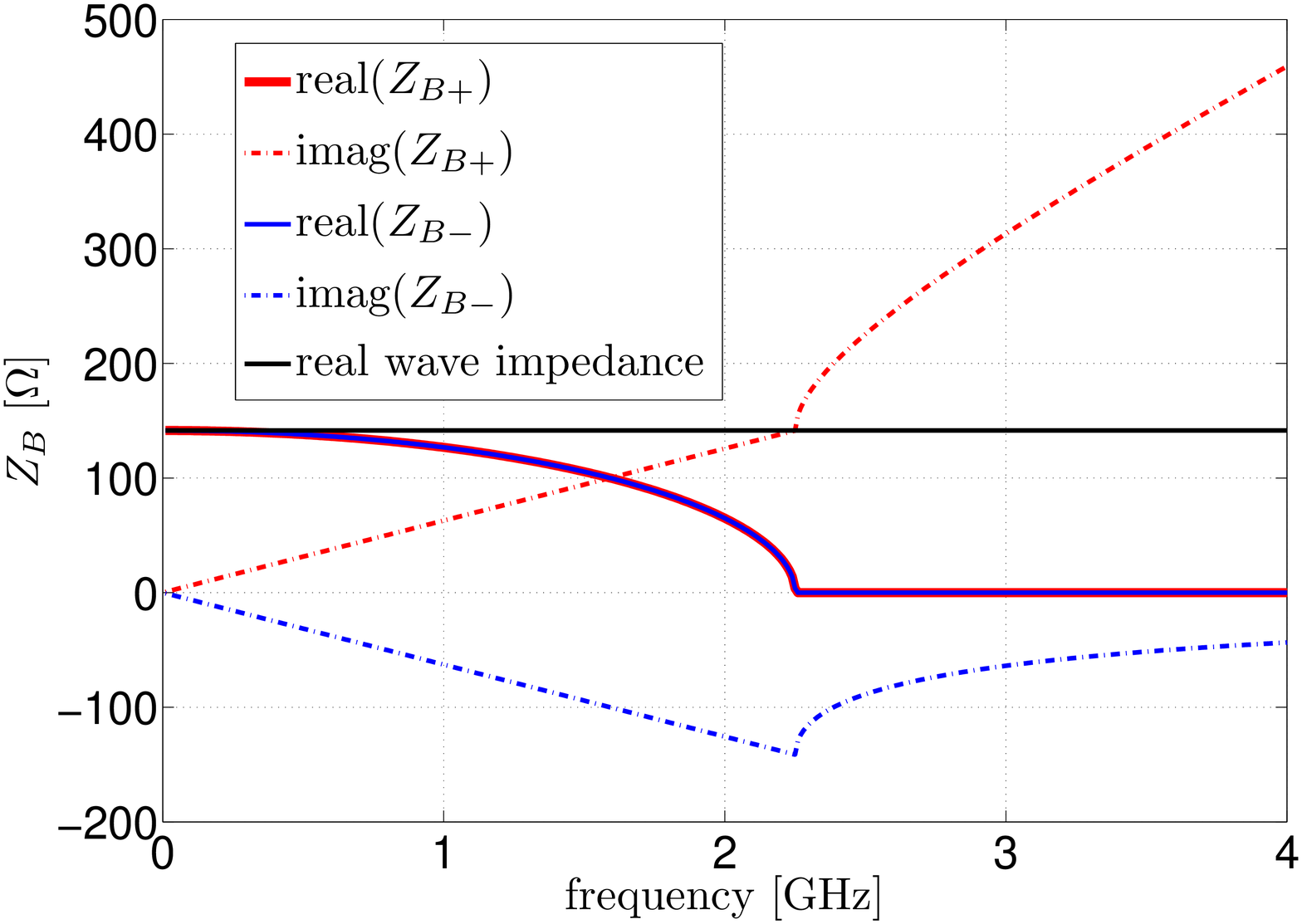,width=0.7\textwidth}
\caption{Bloch impedance for the asymmetric unit cell of Fig.~\ref{fig:comparisona} with $Z = j \omega L$ and $Y = j \omega C$ having component values $L = 20$~nH and $C = 1$~pF.} \label{fig:LC_Bloch}
\end{figure}

\begin{figure}[!ht]
\centering \epsfig{file=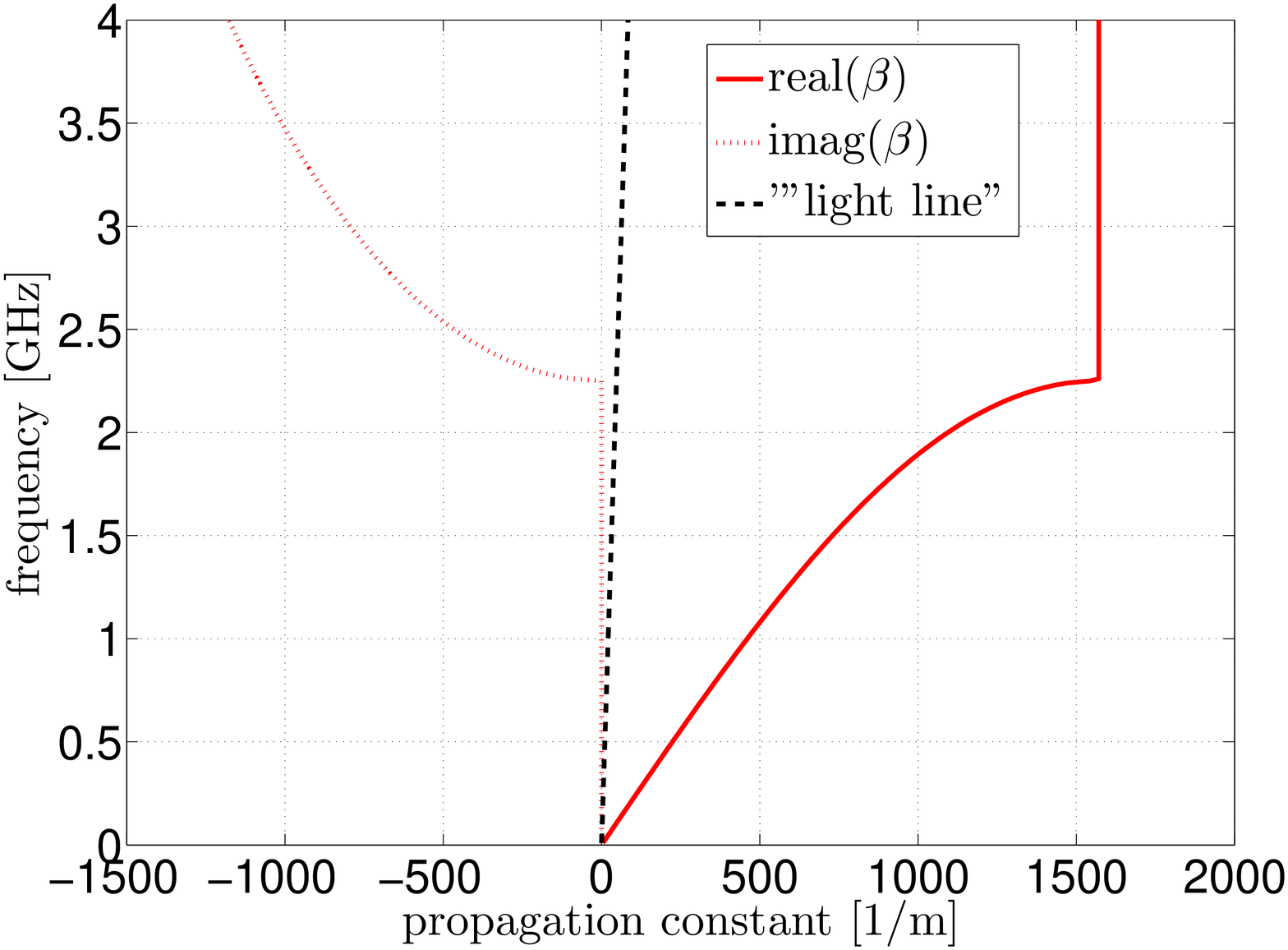,width=0.7\textwidth}
\caption{Dispersion in an infinite cascade for both unit cells shown in
Fig.~\ref{fig:comparison} with $Z = j \omega L$ and $Y = j \omega C$ having component values $L = 20$~nH and $C = 1$~pF.} \label{fig:LC_prop}
\end{figure}

\begin{figure}[!ht]
\centering \epsfig{file=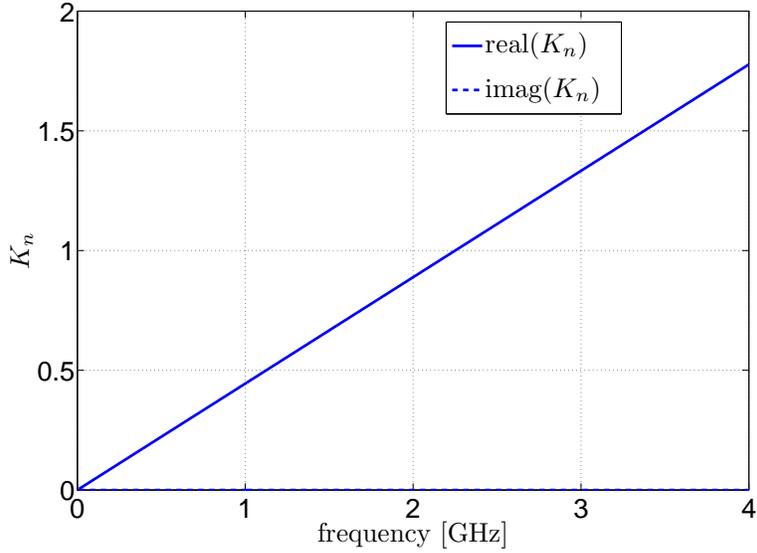,width=0.7\textwidth}
\caption{Normalized omega coefficient for the asymmetric unit cell of
Fig.~\ref{fig:comparisona} with $Z = j \omega L$ and $Y = j \omega C$ having component values $L = 20$~nH and $C = 1$~pF.} \label{fig:LC_Kn}
\end{figure}

\begin{figure}[!ht]
\centering \epsfig{file=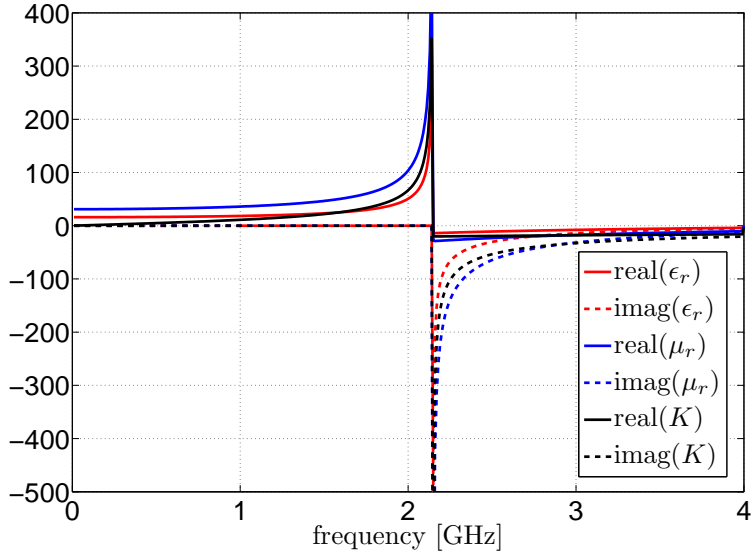,width=0.7\textwidth}
\caption{Permittivity, permeability, and omega coefficient for the asymmetric unit cell of Fig.~\ref{fig:comparisona}
with $Z = j \omega L$ and $Y = j \omega C$ having component values $L = 20$~nH and $C = 1$~pF.} \label{fig:LC_epsmu}
\end{figure}

Let us first consider the asymmetric unit cell shown in
Fig.~\ref{fig:comparisona} with $Z = j \omega L$ and $Y = j \omega
C$ having the component values $L = 20$~nH and $C = 1$~pF. We also
add some small losses in the series branch ($R=0.001$~$\Omega$). This is done just to avoid numerical problems in calculations and does not change the general conclusions. The period of the structure is 2~mm. The effect of the electrically
short TL segments is neglected, as before. Let us first take a look
at the Bloch impedance and dispersion in an infinite cascade of such
unit cells shown in Figs.~\ref{fig:LC_Bloch} and \ref{fig:LC_prop}.
As can be expected, the circuit has low-pass behavior. In the
passband, the Bloch impedance is complex. The real part of the Bloch
impedance is the same for both propagation directions while the
imaginary part has the same magnitude but different signs. In the
stopband, the real part of the Bloch impedance is zero for both
propagation direction, but the imaginary part has different signs
and magnitudes. Notably, the wave impedance term corresponding to
$\sqrt{\mu/\epsilon}$ in omega media (denoted with the black line in
Fig.~\ref{fig:LC_Bloch}) is constant from 0 to 4 GHz. The normalized
omega coefficient $K_n$ is shown in Fig.~\ref{fig:LC_Kn}. As can be
expected, as $K_n$ becomes larger than one, there is no propagation.
Finally,  in Fig.~\ref{fig:LC_epsmu} the permittivity, permeability,
and the omega coefficient $K = K_n \sqrt{\epsilon_t \mu_t}$ are
shown. In the passband, permittivity, permeability, and omega
coefficient are real and positive and their values increase as we
get closer to the stop band. They show resonant behavior in between
the passband and the stopband and have negative values in the
stopband.

\begin{figure}[!ht]
\centering \epsfig{file=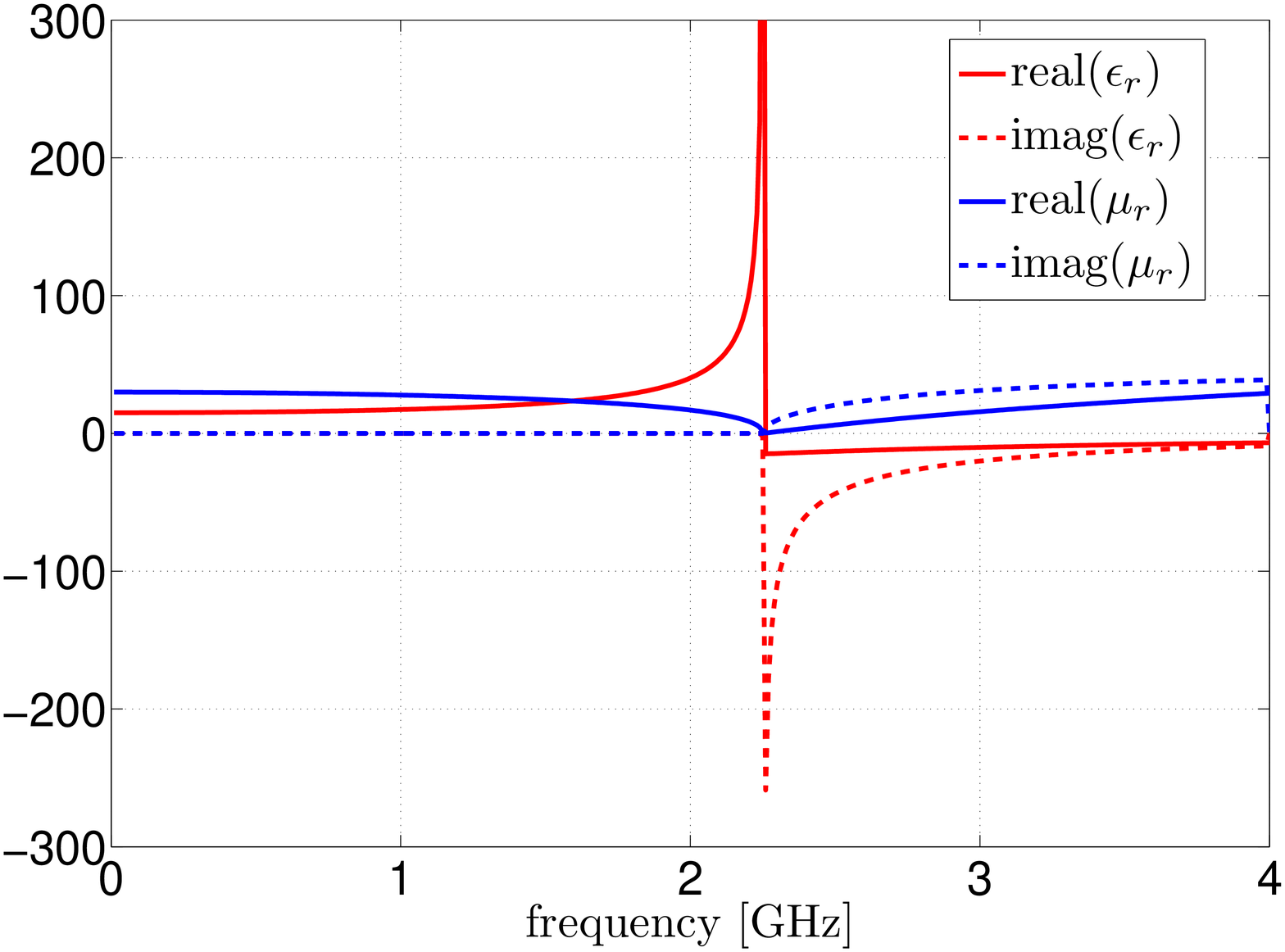,width=0.7\textwidth}
\caption{Permittivity and permeability for the symmetric unit cell of Fig.~\ref{fig:comparisonb}
with $Z = j \omega L/2$ and $Y = j \omega C$ having component values $L = 20$~nH and $C = 1$~pF.} \label{fig:LC2_epsmu}
\end{figure}

Now, let us look at  the symmetric unit cell of
Fig.~\ref{fig:comparisonb} with $Z = j \omega L/2$ and $Y = j \omega
C$ having the component values $L = 20$~nH and $C = 1$~pF. As
before, the period is 2~mm and we have included some small losses in
each series branch ($R/$2=0.0005$~\Omega$). As the period is
electrically small and the effect of the TL segments can therefore
be neglected, this physically corresponds to the earlier case when
an infinite cascade is considered. Obviously, this means that the
dispersion is the same as before. The Bloch impedance, on the other hand, is the same for both propagation directions in this case. The Bloch
impedance is equal to the magnetodielectric wave impedance of
(\ref{eq:mu_over_eps}) and can be written  as $Z_B =
\sqrt{L/C-\omega^2 L^2/4}$. Therefore, the Bloch impedance is
constant with the value $Z_B = \sqrt{L/C}$ only if the product
$\omega^2 L^2/4$ is sufficiently small. Clearly, if we consider a segment of a conventional, truly homogeneous TL with \textit{distributed} series inductance and shunt capacitance, the term $\omega^2 L^2/4$ always approaches zero as the length of the segment approaches zero. Also, if we extract the
material parameters, the permittivity and permeability are notably
different than in the previous case, as can be observed from
Fig.~\ref{fig:LC2_epsmu}. Instead of both permittivity and
permeability increasing in the passband, the permeability decreases.
Typically, when we want to homogenize a periodical structure, we
look at the lower passband where the permittivity and permeability
are weakly dispersive. In this frequency range, the two solutions
are approximately equal, as $|K_n|$ is much smaller than unity in
the asymmetric case. In fact, the negative frequency derivative of
the permeability is not a physical result for passive media as for
passive low-loss media we must have $\frac{\partial
\epsilon(\omega)}{\partial \omega} > 0$ and $\frac{\partial
\mu(\omega)}{\partial \omega} > 0$ ~\cite{Landau}. Also, the
imaginary part of the permeability is positive in the stopband which
also implies that the medium appears to be active. Therefore, this
parameter extraction can only be valid at very low frequencies where
the derivatives are approximately zero. This is not true for the
parameter extraction in the previous asymmetric case, as in that case both
derivatives are positive at all frequencies, and the anti-resonant
artifact of one of the material parameters is removed. Notably, this only happens when the inductance of the other series inductor in the T-circuit is reduced to zero. Even if asymmetry (i.e., omega coupling) is introduced into the full T-circuit by making the inductance of one of the series inductors smaller than the inductance of the other one, the problem of non-physical permittivity and permeability remains as long as both inductors have non-zero values.

\begin{figure}[!ht]
\centering \epsfig{file=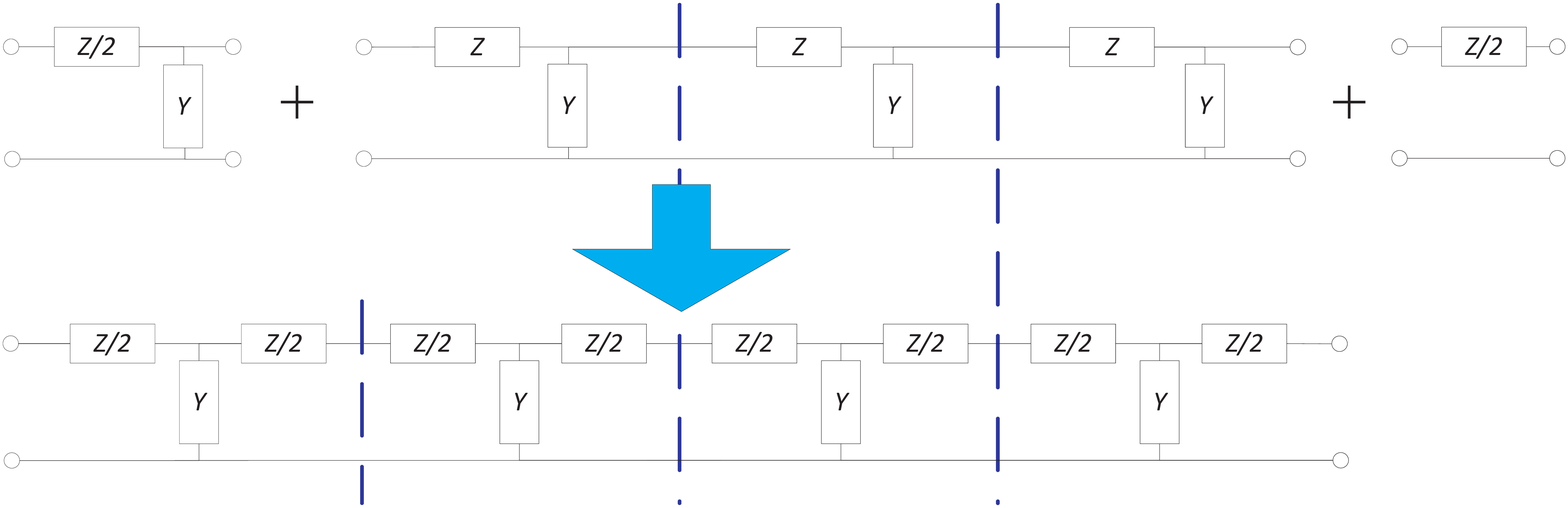,width=0.99\textwidth}
\caption{In a finite sample, topologies of the first
and last unit cells uniquely define the symmetry or asymmetry of the
unit cell. Therefore, a set of $n$ complete asymmetric unit cells can be transformed into $n+1$ complete symmetric unit cells by adding a series element and a parallel element to one end of the cascade and a single series element to the other. } \label{fig:incomplete_uc}
\end{figure}

For an infinite structure, there is no physical reason to select one
unit cell topology over the other. However, as soon as we consider
finite-sized samples, we will see that the topologies of the first
and last unit cells in fact define the symmetry or asymmetry of the
unit cell and in this way also uniquely define the omega coupling
coefficient. If the overall sample is formed by a set of $n$
\emph{complete} unit cells, the effective parameter model with three
effective parameters becomes unique. If at one of the interfaces the
unit cell is incomplete, this can be accounted for by an addition of
a series or parallel circuit element (equivalent to an additional
surface current sheet in the effective medium model) or
alternatively the sample can be interpreted as a set of $n-1$
complete unit cells terminated in an extra circuit element. Moreover, a set of $n$ complete asymmetric unit cells can be transformed into $n+1$ complete symmetric unit cells by adding a series element and a parallel element to one end of the cascade and a single series element to the other. This is illustrated in Fig.~\ref{fig:incomplete_uc}.

\section{Omega parameter in effective medium models of
metamaterials and metasurfaces}

Periodically loaded transmission-line model or a periodical chain of
connected two-port networks is one of the common approaches in
modeling composite materials. Here, the electromagnetic properties
of each period are modeled by the S- or Z-parameters of a two-port
network, connected by short sections of transmission lines or
directly. The main simplification in this model is that the adjacent
layers of a multi-layer structure interact only by the single
fundamental plane-wave harmonic, which is valid when the period
along the propagation direction is larger than the period in the
transverse plane.

For a spatially infinite periodical material a T- or $\Pi$-circuit
can be equally adopted for any passive structure. If the structure
topology is symmetric with respect to the propagation direction (the
simplest example is a 3D periodical array of spherical inclusions),
it is natural to adopt a symmetrical T- or $\Pi$-circuits as models
of each unit cell. As we saw in the previous example, the extracted
effective material parameters (permittivity and permeability) show
non-physical behavior close to the resonant frequency of the unit
cells. This is the well-known artifact of anti-resonant behavior of
one of the two equivalent material parameters of composite materials
containing resonant inclusions (metamaterials), see e.g. Refs.
\onlinecite{A1,A2,A4,A6}. The physical interpretations of this
effect and approaches to developing better and more appropriate
homogenization models have been widely discussed in the literature
\cite{anti1,anti2,anti3,anti4,alu2,alu1,Hol,Alitalo_anti}.
Following these studies, the main conclusions are that anti-resonances
appear because we deal with composites containing resonant particles
of a finite electrical size and the model of two equivalent
parameters extracted from the plane-wave reflection and transmission
coefficients for planar samples is not adequate (additional
effective parameters are needed to account for spatial dispersion in
the structure).

The results of the previous section clearly have important
implications for the problem of composite material homogenization
and defining physically meaningful effective permittivity and
permeability. We have seen that the material parameters extracted
from the symmetrical-unit-cell model exhibit the non-physical
anti-resonant behavior while the parameters given  by the
asymmetrical-unit-cell model \emph{of the same actual structure}
show the physical Lorentz resonances, as is expected. From the point of view of extracted parameters, this means
that one can get non-physical anti-resonant permeability and zero
omega coupling coefficient (symmetric cells) or physical
permittivity and permeability accompanied by positive or negative
omega coupling coefficients. For the
infinite periodical structure the definition of the unit cell is
arbitrary: one can define a symmetric or different asymmetric cells
at will. For an infinitely periodic structure, there is no
physical reason to select one of these models. However, for a finite
structure the topology of the unit cells at the boundaries defines
the appropriate \emph{bulk} effective material model. This is in
line with the Ewald-Oseen extinction theorem (e.g., Ref. \onlinecite{Born}),
which states that the incident field is compensated and the
refracted wave is created by molecules (inclusions) located at the
interface of a medium sample. Previously, it was suggested to
introduce additional surface parameters (additional surface currents)
on the interfaces of a composite slab to remove the anti-resonant
artifact \cite{anti4,Hol,Hol1,Hol2}, but in these models the
bulk properties have been modeled by two parameters, without any
bi-anisotropy. On the other hand, additional bi-anisotropy parameter
has been introduced in Refs. \onlinecite{alu2,alu1,Alu_new}, however, in that
model the field-coupling terms explicitly depend on the propagation
constants of partial plane waves in the sample, which makes the
model difficult to use for samples of a finite thickness. Our
present results give a simple and physically meaningful effective
medium model where the first-order spatial dispersion effects due to
electrically finite size of unit cells are accounted for by an
additional omega coupling parameter. This model does not suffer from
non-physical anti-resonant artifacts in extracted permittivity or
permeability and correctly describes asymmetries in reflection from
different sides of composite slabs. For extraction of all three
effective parameters from reflection and transmission data,
reflection should be measured from both sides of the sample slab.
Finally we note that most of the known experimental samples of
metasurfaces and metamaterials indeed have the topology
corresponding to bi-anisotropic omega coupling (most commonly,
samples are formed by dielectric substrate layers with various metal
patterns printed on \emph{one} side of the substrate). For such
metamaterial layers proper introduction of the omega parameter
removes the anti-resonant artifacts in the effective permittivity
and permeability. Note that even if the individual inclusions in the
bulk are not bi-anisotropic, the bi-anisotropy of the surface (the
topology of the first layer of particles) defines bi-anisotropic
response of the whole sample, seen as a bulk bi-anisotropy
coefficient, as we saw in the above analysis of periodically loaded
transmission lines.

\section{Realization using unit cells with coupled inductors}
\begin{figure}[!ht]
\centering \epsfig{file=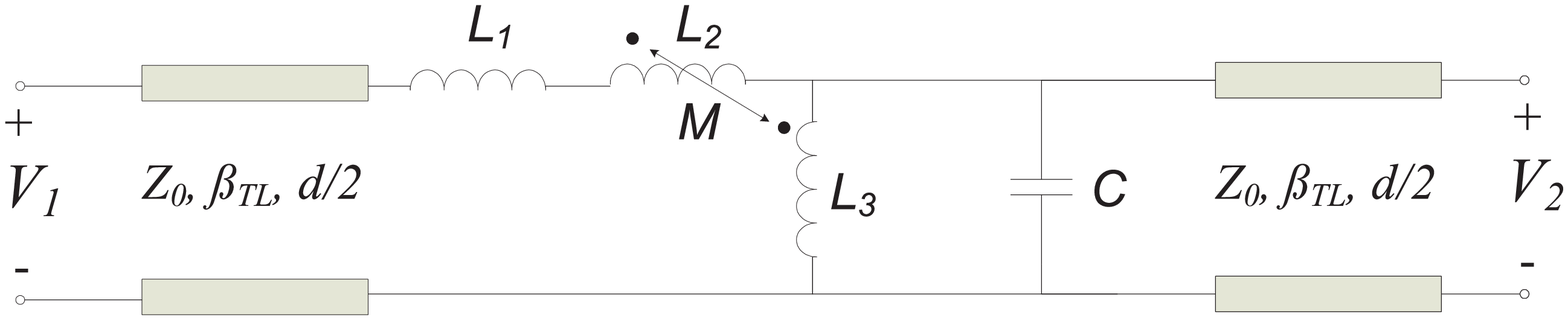,width=0.7\textwidth}
\caption{Circuit topology.} \label{fig:coupledLs}
\end{figure}

\begin{figure}[!ht]
\centering \epsfig{file=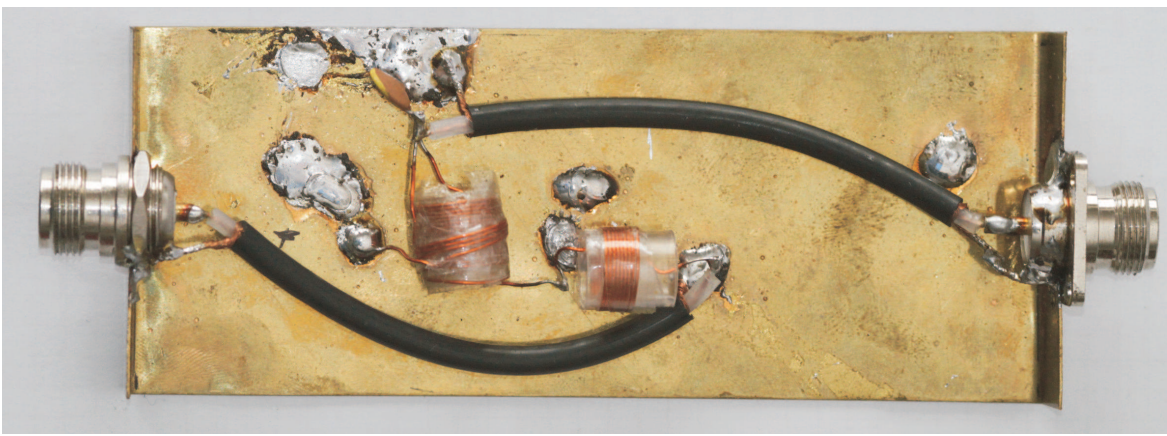, trim = 1cm 12cm 1cm 15cm, clip = true,width=0.95\textwidth}
\caption{Manufactured unit cell.} \label{fig:manufactured}
\end{figure}

As discussed above, LC-loaded TLs have low-pass type of behavior
with the normalized omega coefficient $K_n$ being zero at DC and
increasing with the frequency. However, the permittivity and
permeability appear to have different signs for the passband and the
stopband as well as a strong resonance between the passband and the
stopband. In omega media, the electric and magnetic responses are
coupled. Following from this very basic fact, one might think of another kind of TL with omega-like response (Fig. \ref{fig:coupledLs}). In this TL, the series and shunt branches (i.e., the magnetic and electric fields) are directly coupled via a mutual inductance. The circuit of
Fig.~\ref{fig:coupledLs} with the component values $L_1 = 1.15~\mu
\textnormal{H}$, $L_2 =3.2~\mu \textnormal{H}$, $L_3 = 0.325~\mu
\textnormal{H}$, $M = 0.816~\mu \textnormal{H}$ and $C =
0.68~\textnormal{nF}$ has been studied numerically as well as
experimentally. Since the main goal of this study is to investigate the basic background physics of omega transmission line, the choice of operating frequency range is completely arbitrary. We have chosen a low RF range ($<$~30 MHz) in which the experiments can be performed easily because the  parasitic capacitances are negligible. The period of the unit cell is 0.3~m and the characteristic impedance
of the air-filled host line is 50 $\Omega$. The period of the unit
cell is assumed to be equal to the total length of the two TL
segments. Using the equivalent circuit for coupled inductors (e.g., Ref. \onlinecite{Nilsson}) and knowing the ABCD matrices for the
basic circuit elements, the ABCD parameters can be easily determined
and the circuit analyzed analytically. The manufactured unit cell
prototype can be seen in Fig.~\ref{fig:manufactured}. The number of windings around the
same 12~mm plastic core for the inductors $L_2$ and
$L_3$ was 18 and 4, respectively, and the diameter of the copper wire was 0.5~mm.
The scattering matrix of coupled inductors was measured with Rohde Schwartz ZVA 8 Vector Network Analyzer (VNA). The extraction of the coupling coefficient from the measurement results gave $k=$~0.8. The inductor $L_1$ was also realized as a
coil with 9 windings around a 12~mm plastic core. For the
TL-segments, the 50 $\Omega$ coaxial cable RG-58/U was used. As
RG-58/U is polyethylene-filled ($\epsilon_r$ = 2.26) instead of
air-filled, the physical length of the cable sections in the measurement
setup was 10~cm. The S-parameters of the assembled the unit cell were measured using the VNA
and converted into ABCD parameters. These parameters were then used to calculate
the Bloch impedances, dispersion in an infinite cascade of such unit
cells as well as the equivalent material parameters.

\begin{figure}[!ht]
  \centering
  \subfloat{\label{fig:Silvios_ZB1}\includegraphics[width=0.49\textwidth]{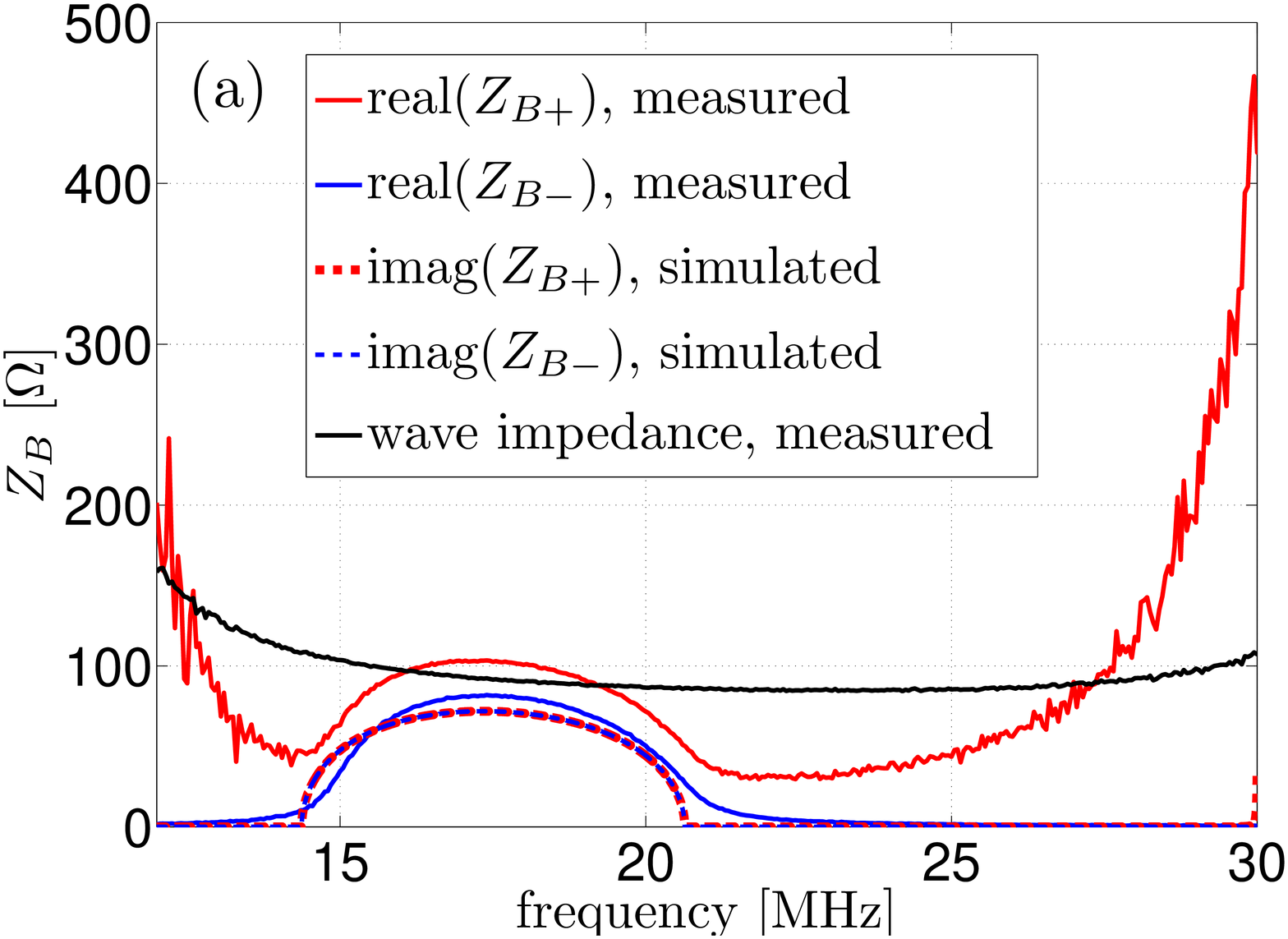}}
  \subfloat{\label{fig:Silvios_ZB2}\includegraphics[width=0.49\textwidth]{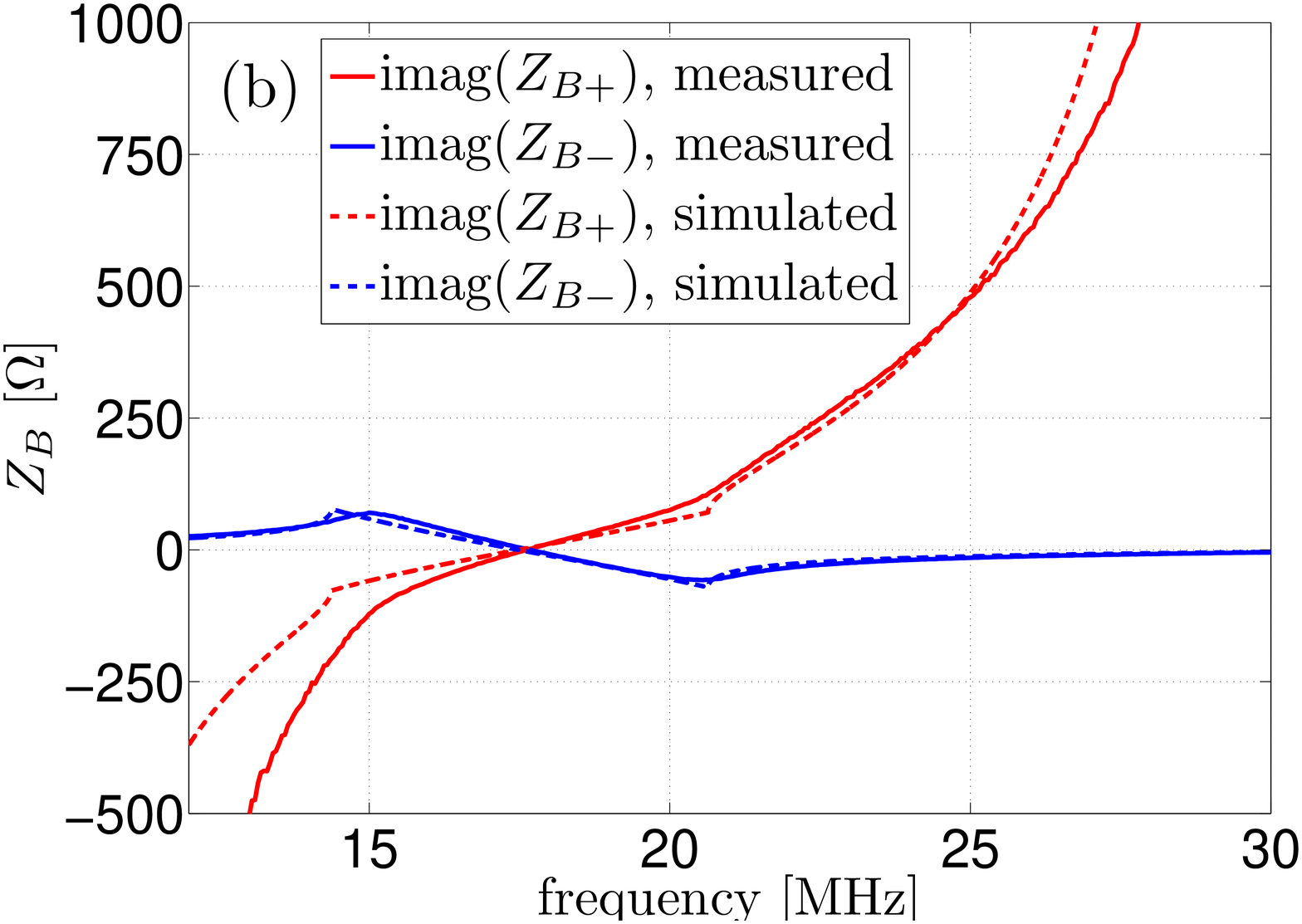}}
  \caption{Bloch impedances for (a) positively and (b) negatively propagating wave for the unit cell of Fig.~\ref{fig:coupledLs}.}
  \label{fig:Silvios_ZB}
\end{figure}

\begin{figure}[!ht]
\centering \epsfig{file=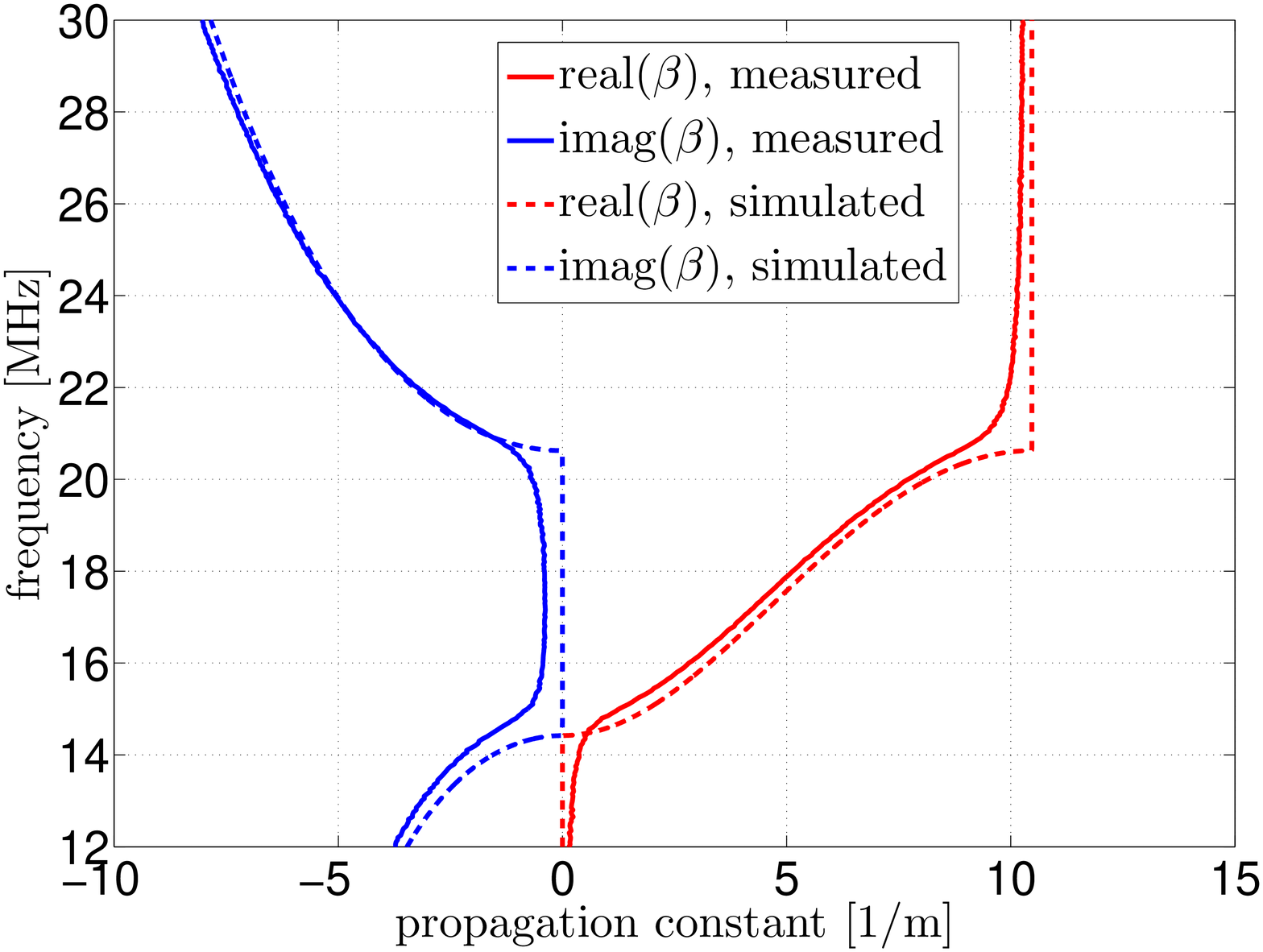,width=0.7\textwidth}
\caption{Dispersion in an infinite cascade for the unit cell of Fig.~\ref{fig:coupledLs}.} \label{fig:Silvios_dispersion}
\end{figure}

\begin{figure}[!ht]
\centering \epsfig{file=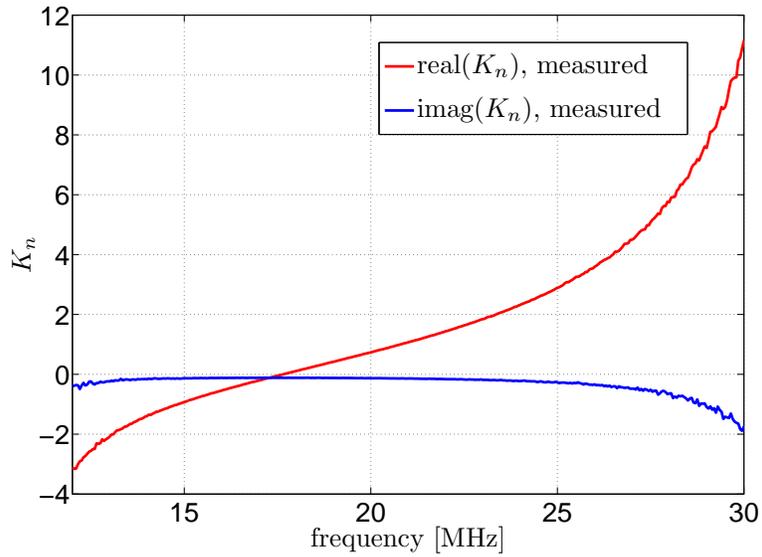,width=0.7\textwidth}
\caption{Normalized omega coefficient extracted from measurement results for the unit cell of Fig.~\ref{fig:coupledLs}.} \label{fig:Silvios_Kn}
\end{figure}

\begin{figure}[!ht]
\centering \epsfig{file=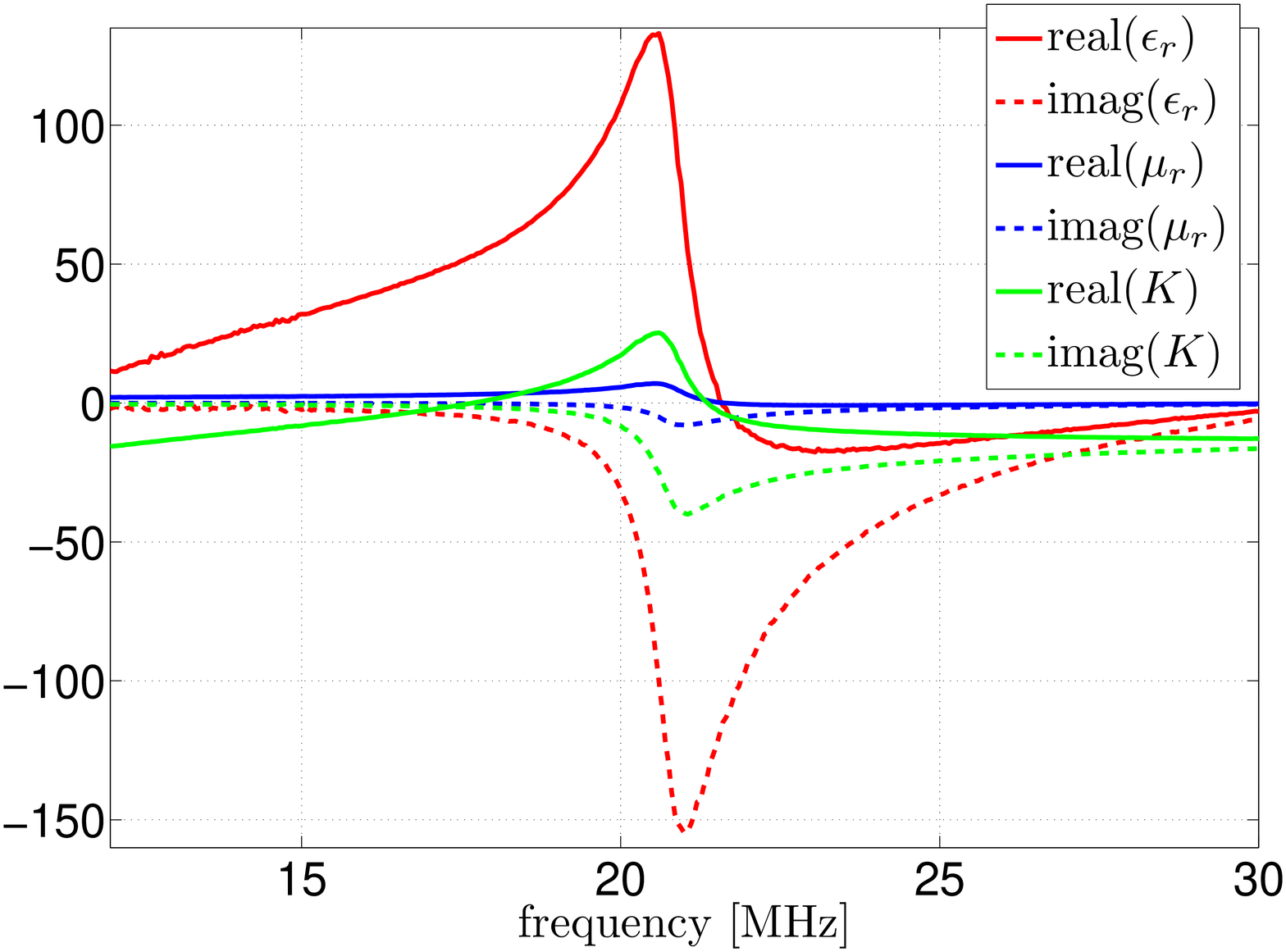,width=0.7\textwidth}
\caption{Permittivity, permeability, and omega coefficient for the unit cell of Fig.~\ref{fig:coupledLs}.} \label{fig:Silvios_matpars}
\end{figure}

The Bloch impedance and the dispersion in an infinite cascade,  in
the frequency range 12~MHz -- 30~MHz calculated both analytically
and based on the measured S-parameters are shown in Fig.~\ref{fig:Silvios_ZB} and Fig.~\ref{fig:Silvios_dispersion},
respectively. Clearly, the TL has band-pass behavior. In the passband, the imaginary part of the Bloch impedance is different for
different propagation directions except for the center of the passband where it is zero. Due to the intrinsic, asymmetric losses in
the coupled inductors, the real part of the Bloch impedance is
different for different propagation directions in the passband and
has a non-zero value in the stopband. The agreement between the
measurement results and the numerical results is in general quite
good. The magnetodielectric wave impedance is also plotted in
Fig.~\ref{fig:Silvios_ZB1}. This can be seen to be fairly constant
in the studied frequency range. The normalized omega coefficient
extracted from the measurement results is shown in
Fig.~\ref{fig:Silvios_Kn}. In the studied frequency range, $K_n$
increases from $-$5 to 11 with the passband appearing when we have
$|K_n|<1$. Finally, the material parameters (permittivity,
permeability, and the omega coefficient) extracted from the
measurements are plotted in Fig.~\ref{fig:Silvios_matpars}.
Permittivity and permeability have similar behavior as in the
earlier LC-loaded case, that is, they have a strong resonance
between the passband and the stopband (or in this case the second
stopband). However, such resonance does not appear between the
first stopband and the passband. The permittivity is, here, notably
larger than the permeability. The difference between permittivity
and permeability could be decreased by increasing the size of the
unit cell. The sign of the omega coupling coefficient changes in the middle of the passband (at 17.7~MHz) while permittivity and permeability are always positive in the studied example. However, by changing the element values appropriately epsilon-negative media (ENG) with omega coupling could also be realized. This gives us more freedom in choosing the material parameters compared to the LC-circuit loading where the permittivity and permeability have the same sign at all frequencies while omega coupling coefficient always has either the same sign (as in the studied example) or the opposite sign compared to permittivity and permeability depending on the order of the two elements within the unit cell. Still, effective permeability can only have positive values in this case due to the absence of capacitive series elements.


%

\section{Comparison to omega wire media}

The most common and the most studied way to realize a medium with omega coupling is to embed metallic wire inclusions having the shape of capital omega letter into a dielectric. The response of this so-called omega wire medium was studied numerically as well as experimentally in Ref.~\onlinecite{Kharina}. In omega wire medium, all the material parameters (permittivity, permeability and omega coefficient) are strongly resonant at the same frequency due to the resonant nature of the particle itself. The omega coefficient is non-zero only very near the resonance frequency of the particle (typically $\lambda$/2 length resonance) and elsewhere the particle is practically invisible to the incident field. Also, the losses near the resonance are considerable with the imaginary parts of permittivity, permeability and omega coupling coefficient being comparable to the corresponding real parts. This is in stark contrast to the presented TL implementations where the omega coupling coefficient is zero only at one frequency (DC for the LC loading circuit and an arbitrary frequency for the circuit utilizing coupled inductors). This is the frequency, where the unit cell appears symmetric. Furthermore, while the losses are large near the resonance also in this case, at frequencies farther from the resonance (where the omega coupling coefficient can still be considerable), the losses are small.

\begin{figure}[!ht]
\centering \epsfig{file=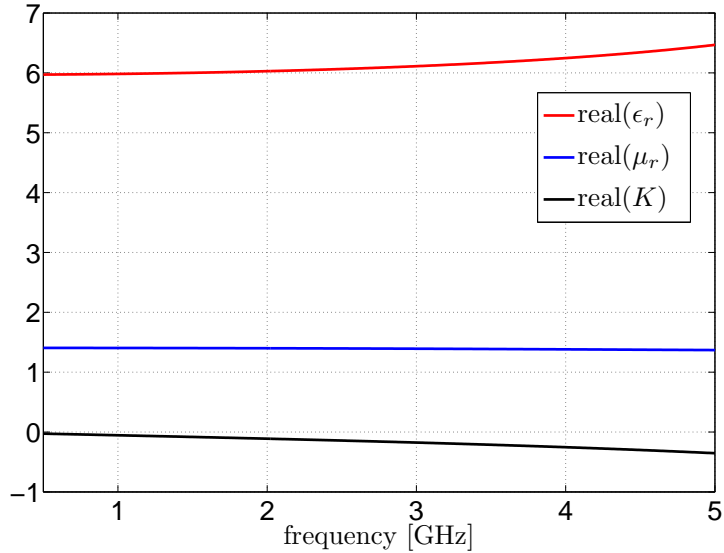,width=0.7\textwidth}
\caption{Permittivity, permeability and omega coefficient for the unit cell of Fig.~\ref{fig:unitcell2} with $Z_1 = 0~\Omega$, $Y_2 = j \omega C$ with $C = 0.11$~pF, $Z_3 = j \omega L$ with $L = 2.1$~nH, $Z_0 = 220~\Omega$, effective permittivity of the TL = 4.8, and $d$ =7~mm.} \label{fig:epsmuK}
\end{figure}

Though it was not the intention here to replicate
the lossy narrow-bandwidth response of omega wire medium, we can
easily realize the effective material parameters of wire omega
medium at a single frequency. For example, the omega wire medium in
Ref.~\onlinecite{Kharina}, consisting of thin wire $\Omega$-shaped
inclusions resonant at 3.5~GHz embedded in epoxy with the resonance
frequency of the medium being 3.2~GHz, has the effective material
parameters $\epsilon_r = 6$, $\mu_r =1.35$, and $K =-0.2$ at 3.1~GHz
as can be read from figures 3,4, and 5 of that paper. The losses are
still fairly small at this frequency and are ignored here. The
simplest way to realize this is to use T-circuit loading as shown in
Fig.~\ref{fig:unitcell2}. The aforementioned material parameters can
be replicated at 3.1~GHz by choosing, for example, the following
values for the unit cell parameters: $Z_1 = 0~\Omega$, $Y_2 = j
\omega C$ with $C = 0.11$~pF, $Z_3 = j \omega L$ with $L = 2.1$~nH,
$Z_0 = 220~\Omega$, effective permittivity of the TL = 4.8, and $d$
= 7~mm. The calculated permittivity, permeability and omega
coefficient for medium characterized by such unit cell are shown in
Fig.~\ref{fig:epsmuK}. Clearly, at 3.1~GHz the material parameters
have the same values as in the example of Ref.~\onlinecite{Kharina}.
However, as can be expected, the dispersion is completely different.
Now, there is very little dispersion in the material parameters
compared to the omega wire medium since we are, in this case,
operating far from any resonance. Obviously, such response at
3.1~GHz could also be replicated with the unit cell utilizing
coupled inductors though the dispersion would again be different
compared to both omega wire media and the T-circuit loaded TL
implementation.

\section{Conclusion}
We have discussed the connection between omega  media and
periodically loaded transmission lines. The effective omega material parameters for
a general periodically loaded TL unit cell have been derived by
comparing the characteristic impedances of omega media and a
periodically loaded TL. Two examples of unit cells providing
omega-like response have been analyzed. We have suggested to model
periodical metamaterials with an equivalent omega medium or omega
transmission lines. This model is free from the anti-resonance in
one of the equivalent material parameters, which becomes possible
because the first-order spatial dispersion effects are properly
accounted for by the omega parameter, instead of being ``embedded''
into the conventional permittivity and permeability. It was shown
that the equivalent omega parameter is determined by the topology of
the unit cells at the boundary of composite-material samples.
Finally, we have numerically and experimentally demonstrated unit
cells of omega transmission lines and compared them to previously proposed realizations based on resonant particles.

\section*{Acknowledgments}
Useful discussions with prof. C. Simovski are gratefully
acknowledged. The authors thank Mr. D. Petricevic for help in
manufacturing the prototype.

\end{document}